\documentclass[12pt]{iopart}
\usepackage{iopams}  
\begin{document}

\title[Parameters and characteristics of the confining 
SU(3)-gluonic field in $\eta^\prime$-meson]
{Estimates for parameters and characteristics of the confining 
SU(3)-gluonic field in an $\eta^\prime$-meson}

\author{Yu P Goncharov}

\address{Theoretical Group, Experimental Physics Department, State 
Polytechnical University, Sankt-Petersburg 195251, Russia}
\ead{ygonch@chern.hop.stu.neva.ru}
\begin{abstract}

The confinement mechanism proposed earlier by the author is applied to estimate 
the possible parameters of the confining SU(3)-gluonic field in 
an $\eta^\prime$-meson. For this aim the electric form factor of an 
$\eta^\prime$-meson is nonperturbatively computed in an explicit analytic form. 
The estimates obtained are also consistent with the width of the electromagnetic 
decay $\eta^\prime\to2\gamma$. The corresponding estimates of the gluon 
concentrations, electric and magnetic colour field strengths are also adduced 
for the mentioned field at the scales of the meson under consideration. 

\end{abstract}

\maketitle

\section{Introduction}
In Refs. \cite{{Gon01},{Gon051},{Gon052}} for the Dirac-Yang-Mills 
system derived from 
QCD-Lagrangian  an unique family of compatible 
nonperturbative solutions was found and explored, which could pretend to 
decsribing confinement of two quarks. 
The applications of the family to the description of both the heavy quarkonia 
spectra \cite{{Gon03},{Gon08a}} and a number of properties of pions, kaons and 
$\eta$-meson \cite{{Gon06},{Gon07a},{Gon07b}} showed that the confinement 
mechanism is qualitatively the same for both light mesons and heavy quarkonia.
At this moment it can be decribed in the following way.

The following main physical reasons underlie linear confinement in the 
mechanism under discussion. The first one is that gluon exchange between 
quarks is realized with the propagator different from the photon one, and 
existence and form of such a propagator is a {\em direct} consequence of the 
unique confining 
nonperturbative solutions of the Yang-Mills equations 
\cite{{Gon051},{Gon052}}. The second reason is that, 
owing to the structure of the mentioned propagator, quarks mainly emit and 
interchange the soft gluons so the gluon condensate (a classical gluon field) 
between quarks basically consists of soft gluons (for more details 
see Refs. \cite{{Gon051},{Gon052}}) but, because of the fact that any gluon 
also emits gluons (still softer), the corresponding gluon concentrations 
rapidly become huge and form a linear confining magnetic colour field of 
enormous strengths, which leads to confinement of quarks. This is by virtue of 
the fact that just the magnetic part of the mentioned propagator is responsible 
for a larger portion of gluon concentrations at large distances since the 
magnetic part has stronger infrared singularities than the electric one. 
In the circumstances 
physically nonlinearity of the Yang-Mills equations effectively vanishes so the 
latter possess the unique nonperturbative confining solutions of the 
Abelian-like form (with the values in Cartan subalgebra of SU(3)-Lie algebra) 
\cite{{Gon051},{Gon052}} which describe 
the gluon condensate under consideration. Moreover, since the overwhelming 
majority of gluons is soft they cannot leave the hadron (meson) until some 
gluons obtain additional energy (due to an external reason) to rush out. So 
we also deal with the confinement of gluons.  

The approach under discussion equips us with the explicit wave functions 
for every two quarks (meson or quarkonium). The wave functions are parametrized 
by a set of 
real constants $a_j, b_j, B_j$ describing the mentioned 
{\em nonperturbative} confining SU(3)-gluonic field (the gluon condensate) and 
they are {\em nonperturbative} modulo square integrable 
solutions of the Dirac equation in the above confining SU(3)-field and also  
depend on $\mu_0$, the reduced
mass of the current masses of quarks forming meson. It is clear that under the 
given approach just constants $a_j, b_j, B_j,\mu_0$ determine all properties 
of any meson (quarkonium), i. e.,  the approach directly appeals to quark 
and gluonic degrees of freedom as should be according to the first principles 
of QCD. Also it is clear that the constants mentioned should be extracted from 
experimental data. 

Such a program has been to a certain extent advanced in 
Refs. \cite{{Gon03},{Gon06},{Gon07a},{Gon07b},{Gon08a}}. The aim of the present 
paper is to continue obtaining estimates for $a_j, b_j, B_j$ for concrete 
mesons starting from experimental data on spectroscopy of one or another meson. 
We here consider an $\eta^\prime$-meson and its electromagnetic decay 
$\eta^\prime\to2\gamma$. 

Of course, when conducting our considerations 
we shall rely on the standard quark model (SQM) based on SU(3)-flavor symmetry 
(see, e. g., Ref. \cite{pdg}), so in accordance with SQM 
$\eta^\prime=\sqrt{1/3}(\bar{u}u+\bar{d}d+\bar{s}s)$ is a superposition of 
three quarkonia; consequently, we shall have three sets of parameters 
$a_j, b_j, B_j$.

Section 2 contains main relations underlying 
description of any mesons (quarkonia) in our 
approach. Section 3 is 
devoted to computing the electric form factor, the root-mean-square radius 
$<r>$ and the magnetic moment of the meson under consideration in an explicit 
analytic form. Section 4 gives an independent estimate for $<r>$ which is used 
in Section 5 for obtaining estimates for parameters of the confining 
SU(3)-gluonic field for an $\eta^\prime$-meson. Also Section 5 contains 
a discussion about whether the obtained 
estimates might also be consistent with the width of 2-photon decay 
$\eta^\prime\to2\gamma$. Section 6 employs the obtained parameters of 
SU(3)-gluonic field to get the corresponding estimates for such characteristics 
of the mentioned field as gluon concentrations, electric and magnetic colour 
field strengths at the scales of an $\eta^\prime$-meson while Section 7 is 
devoted to discussion and concluding remarks. 
                      
At last, Appendices $A$ and $B$ contain the detailed description 
of main building blocks for meson wave functions in the approach under 
discussion, respectively: eigenspinors of the Euclidean Dirac operator on 
2-sphere ${\mathbb S}^2$ and radial parts for the modulo square integrable 
solutions of Dirac equation in the confining SU(3)-Yang-Mills field. 
 
Further we shall deal with the metric of
the flat Minkowski spacetime $M$ that
we write down (using the ordinary set of local spherical coordinates
$r,\vartheta,\varphi$ for the spatial part) in the form
$$ds^2=g_{\mu\nu}dx^\mu\otimes dx^\nu\equiv
dt^2-dr^2-r^2(d\vartheta^2+\sin^2\vartheta d\varphi^2)\>. \eqno(1)$$
Besides, we have $|\delta|=|\det(g_{\mu\nu})|=(r^2\sin\vartheta)^2$
and $0\leq r<\infty$, $0\leq\vartheta<\pi$,
$0\leq\varphi<2\pi$.

Throughout the paper we employ the Heaviside-Lorentz system of units 
with $\hbar=c=1$, unless explicitly stated otherwise, so the gauge coupling 
constant $g$ and the strong coupling constant ${\alpha_s}$ are connected by 
the relation $g^2/(4\pi)=\alpha_s$. 
Further, we shall denote by $L_2(F)$ the set of the modulo square integrable
complex functions on any manifold $F$ furnished with an integration measure, 
then $L^n_2(F)$ will be the $n$-fold direct product of $L_2(F)$
endowed with the obvious scalar product while $\dag$ and $\ast$ stand, 
respectively, for Hermitian and complex conjugation. Our choice of Dirac 
$\gamma$-matrices conforms to the so-called standard representation and is 
the same as in Ref. \cite{Gon06}. At last $\otimes$ means 
tensorial product of matrices and $I_n$ is the unit $n\times n$ matrix so that, 
e.g., we have 
$$I_3\otimes\gamma^\mu=
\pmatrix{\gamma^\mu&0&0\cr 0&\gamma^\mu&0\cr 0&0&\gamma^\mu\cr}$$ 
for any Dirac $\gamma$-matrix $\gamma^\mu$ and so forth. 

When calculating we apply the 
relations $1\ {\rm GeV^{-1}}\approx0.1973269679\ {\rm fm}\>$,
$1\ {\rm s^{-1}}\approx0.658211915\times10^{-24}\ {\rm GeV}\>$, 
$1\ {\rm V/m}\approx0.2309956375\times 10^{-23}\ {\rm GeV}^2$, 
$1\ {\rm T}=4\pi\times10^{-7} {\rm H/m}\times1\ {\rm A/m}
\approx0.6925075988\times 10^{-15}\ {\rm GeV}^2 $. 

Finally, for the necessary estimates we shall employ the $T_{00}$-component 
(volumetric energy density ) of the energy-momentum tensor for a 
SU(3)-Yang-Mills field which should be written in the chosen system of units 
in the form
$$T_{\mu\nu}=-F^a_{\mu\alpha}\,F^a_{\nu\beta}\,g^{\alpha\beta}+
{1\over4}F^a_{\beta\gamma}\,F^a_{\alpha\delta}g^{\alpha\beta}g^{\gamma\delta}
g_{\mu\nu}\>. \eqno(2) $$

\section{Main relations}
As was mentioned above, our considerations shall be based on the unique family 
of compatible nonperturbative solutions for 
the Dirac-Yang-Mills system (derived from QCD-Lagrangian) studied at the whole 
length in Refs. \cite{{Gon01},{Gon051},{Gon052}}.  Referring for more details 
to those references, let us briefly decribe and specify only the relations 
necessary to us in the present paper. 

One part of the mentioned family is presented by the unique nonperturbative 
confining solution of the SU(3)-Yang-Mills 
equations for the gluonic field $A=A_\mu dx^\mu=
A^a_\mu \lambda_adx^\mu$ ($\lambda_a$ are the 
known Gell-Mann matrices, $\mu=t,r,\vartheta,\varphi$, $a=1,...,8$) and looks 
as follows 
$$ {\cal A}_{1t}\equiv A^3_t+\frac{1}{\sqrt{3}}A^8_t =-\frac{a_1}{r}+A_1 \>,
{\cal A}_{2t}\equiv -A^3_t+\frac{1}{\sqrt{3}}A^8_t=-\frac{a_2}{r}+A_2\>,$$
$${\cal A}_{3t}\equiv-\frac{2}{\sqrt{3}}A^8_t=\frac{a_1+a_2}{r}-(A_1+A_2)\>, $$
$$ {\cal A}_{1\varphi}\equiv A^3_\varphi+\frac{1}{\sqrt{3}}A^8_\varphi=
b_1r+B_1 \>,
{\cal A}_{2\varphi}\equiv -A^3_\varphi+\frac{1}{\sqrt{3}}A^8_\varphi=
b_2r+B_2\>,$$
$${\cal A}_{3\varphi}\equiv-\frac{2}{\sqrt{3}}A^8_\varphi=
-(b_1+b_2)r-(B_1+B_2)\> \eqno(3)$$
with the real constants $a_j, A_j, b_j, B_j$ parametrizing the family. 
The word {\em unique} should be understood in the strict mathematical sense. 
In fact in Ref. \cite{Gon051} the following theorem was proved:

{\em The unique exact spherically symmetric (nonperturbative) solutions (i.e. 
depending only on $r$) of SU(3)-Yang-Mills equations in Minkowski spacetime 
consist of the family of (3)}.

It should be noted that solution (3) was found early in 
Ref. \cite{Gon01} but its uniqueness was proved just in Ref. \cite{Gon051} 
(see also Ref. \cite{Gon052}). Besides, in Ref. \cite{Gon051} (see also 
Ref. \cite{Gon06}) it was shown that the above unique confining solutions (3) 
satisfy the so-called Wilson confinement criterion \cite{Wil}. Up to now 
nobody contested this result so if we want to describe interaction between 
quarks by spherically symmetric SU(3)-fields then they can be only those  
from the above theorem.

As has been repeatedly explained in 
Refs. \cite{{Gon051},{Gon052},{Gon03},{Gon06}}, parameters $A_{1,2}$ of 
solution (3) are inessential for physics in question and we can 
consider $A_1=A_2=0$. Obviously we have 
$\sum_{j=1}^{3}{\cal A}_{jt}=\sum_{j=1}^{3}{\cal A}_{j\varphi}=0$ which 
reflects the fact that for any matrix 
${\cal T}$ from SU(3)-Lie algebra we have ${\rm Tr}\,{\cal T}=0$. 
Also, as has been repeatedly discussed by us earlier (see, e. g., 
Refs. \cite{{Gon051},{Gon052}}), from the above form it is clear that 
the solution (3) is a configuration describing the electric Coulomb-like colour 
field (components $A^{3,8}_t$) and the magnetic colour field linear in $r$ 
(components $A^{3,8}_\varphi$) and we wrote down
the solution (3) in the combinations that are just 
needed further to insert into the Dirac equation (4). 

Another part of the family is given by the unique nonperturbative modulo 
square integrable solutions of the Dirac equation in the confining 
SU(3)-field of (3) $\Psi=(\Psi_1, \Psi_2, \Psi_3)$ 
with the four-dimensional Dirac spinors 
$\Psi_j$ representing the $j$th colour component of the meson, 
so $\Psi$ may describe the relative motion (relativistic bound states) of two 
quarks in mesons and the mentioned Dirac equation is written as follows 
$$i\partial_t\Psi\equiv  
i\pmatrix{\partial_t\Psi_1\cr \partial_t\Psi_2\cr \partial_t\Psi_3\cr}=
H\Psi\equiv\pmatrix{H_1&0&0\cr 0&H_2&0\cr 0&0&H_3\cr}
\pmatrix{\Psi_1\cr\Psi_2\cr\Psi_3\cr}=
\pmatrix{H_1\Psi_1\cr H_2\Psi_2\cr H_3\Psi_3\cr}
                   \,,\eqno(4)$$
where the Hamiltonian $H_j$ is 
$$H_j=\gamma^0\left[\mu_0-i\gamma^1\partial_r-i\gamma^2\frac{1}{r}
\left(\partial_\vartheta+\frac{1}{2}\gamma^1\gamma^2\right)-
i\gamma^3\frac{1}{r\sin{\vartheta}}
\left(\partial_\varphi+\frac{1}{2}\sin{\vartheta}\gamma^1\gamma^3
+\frac{1}{2}\cos{\vartheta}\gamma^2\gamma^3\right)\right]$$
$$-g\gamma^0\left(\gamma^0{\cal A}_{jt}+\gamma^3\frac{1}{r\sin{\vartheta}}
{\cal A}_{j\varphi}\right) \eqno(5)  $$                           
with the gauge coupling constant $g$ while $\mu_0$ is a mass parameter and one 
can consider it to be the reduced mass which is equal, {\it e. g.}, for 
quarkonia, to half the current mass of quarks forming a quarkonium.

Then the unique nonperturbative modulo square integrable solutions of (4) 
are (with Pauli matrix $\sigma_1$)  
$$\Psi_j=e^{-i\omega_j t}\psi_j\equiv 
e^{-i\omega_j t}r^{-1}\pmatrix{F_{j1}(r)\Phi_j(\vartheta,\varphi)\cr\
F_{j2}(r)\sigma_1\Phi_j(\vartheta,\varphi)}\>,j=1,2,3\eqno(6)$$
with the 2D eigenspinor $\Phi_j=\pmatrix{\Phi_{j1}\cr\Phi_{j2}}$ of the
Euclidean Dirac operator ${\cal D}_0$ on the unit sphere ${\mathbb S}^2$, while 
the coordinate $r$ stands for the distance between quarks. The explicit form of 
$\Phi_j$ is not needed here and
can be found in Refs. \cite{{Gon052},{Gon99}}. For the purpose of the present 
paper we shall adduce the necessary spinors in Appendix A. Spinors $\Phi_j$ form an 
orthonormal basis in $L_2^2({\mathbb S}^2)$.
 We can call the quantity $\omega_j$ 
the relative energy of the $j$th colour component of a meson (while $\psi_j$ is 
the wave function of a stationary state for the $j$th colour component), but 
we can see that 
if we want to interpret (4) as an equation for eigenvalues of the relative 
motion energy, i. e.,  to rewrite it in the form $H\psi=\omega\psi$ with 
$\psi=(\psi_1, \psi_2, \psi_3)$ then we should put $\omega=\omega_j$ for 
any $j$ so that $H_j\psi_j=\omega_j\psi_j=\omega\psi_j$. In this situation, 
if a meson is composed of quarks $q_{1,2}$ with different flavours then 
the energy spectrum of the meson will be given 
by $\epsilon=m_{q_1}+m_{q_2}+\omega$ with the current quark masses $m_{q_k}$ (
rest energies) of the corresponding quarks. On the other hand for 
determination of $\omega_j$ the following quadratic equation can be obtained 
\cite{{Gon01},{Gon051},{Gon052}}
$$[g^2a_j^2+(n_j+\alpha_j)^2]\omega_j^2-
2(\lambda_j-gB_j)g^2a_jb_j\,\omega_j+
[(\lambda_j-gB_j)^2-(n_j+\alpha_j)^2]g^2b_j^2-
\mu_0^2(n_j+\alpha_j)^2=0\>,  \eqno(7)   $$
which yields 
$$\omega_j=\omega_j(n_j,l_j,\lambda_j)=$$ 
$$\frac{\Lambda_j g^2a_jb_j\pm(n_j+\alpha_j)
\sqrt{(n_j^2+2n_j\alpha_j+\Lambda_j^2)\mu_0^2+g^2b_j^2(n_j^2+2n_j\alpha_j)}}
{n_j^2+2n_j\alpha_j+\Lambda_j^2}\>, j=1,2,3\>,\eqno(8)$$

where $a_3=-(a_1+a_2)$, $b_3=-(b_1+b_2)$, $B_3=-(B_1+B_2)$, 
$\Lambda_j=\lambda_j-gB_j$, $\alpha_j=\sqrt{\Lambda_j^2-g^2a_j^2}$, 
$n_j=0,1,2,...$, while $\lambda_j=\pm(l_j+1)$ are
the eigenvalues of Euclidean Dirac operator ${\cal D}_0$ 
on a unit sphere with $l_j=0,1,2,...$. It should be noted that in the  
papers \cite{{Gon01},{Gon051},{Gon052},{Gon03},{Gon06}} we used the ansatz (6) 
with the factor $e^{i\omega_j t}$ instead of $e^{-i\omega_j t}$ but then the 
Dirac equation (4) would look as $-i\partial_t\Psi= H\Psi$ and in equation (7) 
the second summand would have the plus sign while the first summand in 
numerator of (8) would have the minus sign. In the papers 
\cite{{Gon07a},{Gon07b}} we returned to the conventional form of 
writing Dirac equation and this slightly modified the equations (7)--(8). In 
the given paper we conform to the same prescription as in 
Refs. \cite{{Gon07a},{Gon07b}}. 

In line with the above we should have $\omega=\omega_1=\omega_2=\omega_3$ in 
energy spectrum $\epsilon=m_{q_1}+m_{q_2}+\omega$ for any meson (quarkonium) 
and this at once imposes two conditions on parameters $a_j,b_j,B_j$ when 
choosing some experimental value for $\epsilon$ at the given current quark 
masses $m_{q_1},m_{q_2}$. 

The general form of the radial parts of (6) is considered in Appendix $B$. 
Within the given paper we need only the radial parts of (6) at $n_j=0$ 
(the ground state) that are [see $(B.5)$]  
$$F_{j1}=C_jP_jr^{\alpha_j}e^{-\beta_jr}\left(1-
\frac{gb_j}{\beta_j}\right), P_j=gb_j+\beta_j, $$
$$F_{j2}=iC_jQ_jr^{\alpha_j}e^{-\beta_jr}\left(1+
\frac{gb_j}{\beta_j}\right), Q_j=\mu_0-\omega_j\eqno(9)$$
with $\beta_j=\sqrt{\mu_0^2-\omega_j^2+g^2b_j^2}$, while $C_j$ is determined 
from the normalization condition
$\int_0^\infty(|F_{j1}|^2+|F_{j2}|^2)dr=\frac{1}{3}$. 
Consequently, we shall gain that $\Psi_j\in L_2^{4}({\mathbb R}^3)$ at any 
$t\in{\mathbb R}$ and, as a result,
the solutions of (6) may describe relativistic bound states (mesons) 
with the energy (mass) spectrum $\epsilon$.
\subsection{Nonrelativistic limit}
It is useful to specify the nonrelativistic limit (when 
$c\to\infty$) for spectrum (8). For this one should replace 
$g\to g/\sqrt{\hbar c}$, 
$a_j\to a_j/\sqrt{\hbar c}$, $b_j\to b_j\sqrt{\hbar c}$, 
$B_j\to B_j/\sqrt{\hbar c}$ and, expanding (8) in $z=1/c$, we shall get
$$\omega_j(n_j,l_j,\lambda_j)=$$
$$\pm\mu_0c^2\left[1\mp
\frac{g^2a_j^2}{2\hbar^2(n_j+|\lambda_j|)^2}z^2\right]
+\left[\frac{\lambda_j g^2a_jb_j}{\hbar(n_j+|\lambda_j|)^2}\,
\mp\mu_0\frac{g^3B_ja_j^2f(n_j,\lambda_j)}{\hbar^3(n_j+|\lambda_j|)^{7}}\right]
z\,+O(z^2)\>,\eqno(10)$$
where 
$f(n_j,\lambda_j)=4\lambda_jn_j(n_j^2+\lambda_j^2)+
\frac{|\lambda_j|}{\lambda_j}\left(n_j^{4}+6n_j^2\lambda_j^2+\lambda_j^4
\right)$. 

As is seen from (10), at $c\to\infty$ the contribution of linear magnetic 
colour field (parameters $b_j, B_j$) to the spectrum really vanishes and the 
spectrum in essence becomes the purely nonrelativistic Coulomb one (modulo 
the rest energy). Also it is 
clear that when $n_j\to\infty$, $\omega_j\to\pm\sqrt{\mu_0^2+g^2b_j^2}$. 
At last, one should specify the weak 
coupling limit of (8), i.e., the case $g\to0$. As is not complicated to see 
from (8), $\omega_j\to\pm\mu_0$ when $g\to0$. But then quantities 
$\beta_j=\sqrt{\mu_0^2-\omega_j^2+g^2b_j^2}\to0$ and wave functions of (9) 
cease to be the modulo square integrable ones at $g=0$, i.e., they cease to 
describe relativistic bound states. Accordingly, this means that the equation 
(8) does not make physical meaning at $g=0$. 

We may seemingly use (8) with various combinations of signes ($\pm$) before 
the second summand in numerators of (8) but, due to (10), it is 
reasonable to take all signs equal to plus which is our choice within the 
paper. Besides, 
as is not complicated to see, radial parts in the nonrelativistic limit have 
the behaviour of form $F_{j1},F_{j2}\sim r^{l_j+1}$, which allows one to call 
quantum number $l_j$ angular momentum for the $j$th colour component though 
angular momentum is not conserved in the field (3) \cite{{Gon01},{Gon052}}. So, 
for a meson (quarkonium) under consideration we should put all $l_j=0$. 

\subsection{Eigenspinors with $\lambda=\pm1$}
Finally it should be noted that spectrum (8) is degenerated owing to the  
degeneracy of eigenvalues for the
Euclidean Dirac operator ${\cal D}_0$ on the unit sphere ${\mathbb S}^2$. 
Namely, each eigenvlalue of ${\cal D}_0$ $\lambda =\pm(l+1), l=0,1,2...$, has 
multiplicity $2(l+1)$, so we has $2(l+1)$ eigenspinors orthogonal to each other. 
Ad referendum we need eigenspinors corresponding to $\lambda =\pm1$ ($l=0$) 
so here is their explicit form [see $(A.16)$]
$$\lambda=-1: \Phi=\frac{C}{2}\pmatrix{e^{i\frac{\vartheta}{2}}
\cr e^{-i\frac{\vartheta}{2}}\cr}e^{i\varphi/2},\> {\rm or}\>\>
\Phi=\frac{C}{2}\pmatrix{e^{i\frac{\vartheta}{2}}\cr
-e^{-i\frac{\vartheta}{2}}\cr}e^{-i\varphi/2},$$
$$\lambda=1: \Phi=\frac{C}{2}\pmatrix{e^{-i\frac{\vartheta}{2}}\cr
e^{i\frac{\vartheta}{2}}\cr}e^{i\varphi/2}, \> {\rm or}\>\>
\Phi=\frac{C}{2}\pmatrix{-e^{-i\frac{\vartheta}{2}}\cr
e^{i\frac{\vartheta}{2}}\cr}e^{-i\varphi/2} 
\eqno(11) $$
with the coefficient $C=1/\sqrt{2\pi}$ (for more details, see 
Appendix $A$). 

\section{Electric form factor, the root-mean-square radius and magnetic moment}
Obviously, we should choose a few quantities that are the most important from 
the physical point of view to characterize 
meson under consideration and then we should evaluate the given quantities 
within the framework of our approach. In the circumstances let us settle on 
the ground state energy (mass) of an $\eta^\prime$-meson, the root-mean-square 
radius of it and the magnetic moment. All three magnitudes are essentially 
nonperturbative ones, and can be calculated only by nonperturbative techniques.

Within the present paper we shall use relations (8) at $n_j=0=l_j$ so 
energy (mass) of meson under consideration is given by $\mu=2m_q+\omega$ with 
$\omega=\omega_j(0,0,\lambda_j)$ for any $j=1,2,3$ whereas 
$$\omega=\frac{g^2a_1b_1}{\Lambda_1}+\frac{\alpha_1\mu_0}
{|\Lambda_1|}=\frac{g^2a_2b_2}{\Lambda_2}+\frac{\alpha_2\mu_0}
{|\Lambda_2|}=\frac{g^2a_3b_3}{\Lambda_3}+\frac{\alpha_3\mu_0}
{|\Lambda_3|}=\mu-2m_q
\>\eqno(12)$$
and, as a consequence, the corresponding meson (quarkonium) wave functions of 
(6) are represented by (9) and (11). 
\subsection{Choice of quark masses and the gauge coupling constant}
It is evident for employing the above relations we have to assign some values 
to quark masses and gauge coupling constant $g$. In accordance with 
Ref. \cite{pdg}, at present the current quark masses necessary to us are 
restricted to intervals $1.5\>{\rm MeV}\le m_u\le 3\>\,{\rm MeV}$, 
$3.0\> {\rm MeV}\le m_d\le 7 \> {\rm MeV}$, 
$95\> {\rm MeV}\le m_s\le 120 \> {\rm MeV}$, 
so we take $m_u=(1.5+3)/2\>\,{\rm MeV}=2.25\>\,{\rm MeV}$, 
$m_d=(3+7)/2\>\,{\rm MeV}=5\>\,{\rm MeV}$, 
$m_s=(95+120)/2\>\,{\rm MeV}=107.5\>\,{\rm MeV}$. 
Under the circumstances, the reduced mass $\mu_0$ of (5) will respectively 
take values $m_u/2, m_d/2, m_s/2$. As to 
gauge coupling constant $g=\sqrt{4\pi\alpha_s}$, it should be noted that 
recently some attempts have been made to generalize the standard formula
for $\alpha_s=\alpha_s(Q^2)=12\pi/[(33-2n_f)\ln{(Q^2/\Lambda^2)}]$ ($n_f$ is 
number of quark flavours) holding true at the momentum transfer 
$\sqrt{Q^2}\to\infty$ 
to the whole interval $0\le \sqrt{Q^2}\le\infty$. We shall employ one such a 
generalization used in Refs. \cite{De1}. It is written as follows 
($x=\sqrt{Q^2}$ in GeV) 
$$ \alpha(x)=\frac{12\pi}{(33-2n_f)}\frac{f_1(x)}{\ln{\frac{x^2+f_2(x)}
{\Lambda^2}}} 
\eqno(13) $$
with 
$$f_1(x)=
1+\left(\left(\frac{(1+x)(33-2n_f)}{12}\ln{\frac{m^2}{\Lambda^2}}-1
\right)^{-1}+0.6x^{1.3}\right)^{-1}\>,\>f_2(x)=m^2(1+2.8x^2)^{-2}\>,$$
wherefrom one can conclude that $\alpha_s\to \pi=3.1415...$ when $x\to 0$, 
i. e., $g\to{2\pi}=6.2831...$. We used (13) at $m=1$ GeV, $\Lambda=0.234$ GeV, 
$n_f=3$, $x=m_{\eta^\prime}=957.78$ MeV to obtain $g=3.91476$ necessary for 
our further computations at the mass scale of an $\eta^\prime$-meson. 

\subsection{Electric form factor}
For each meson (quarkonium) with the wave function $\Psi=(\Psi_j)$ of (6) we 
can define the  
electromagnetic current $J^\mu=\overline{\Psi}(I_3\otimes\gamma^\mu)\Psi=
(\Psi^{\dag}\Psi,\Psi^{\dag}(I_3\otimes{\bf \alpha})\Psi)=(\rho,{\bf J})$, 
${\bf \alpha}=\gamma^0{\bf\gamma}$.  
The electric form factor $f(K)$ is the Fourier transform of $\rho$
$$ f(K)= \int\Psi^{\dag}\Psi e^{-i{\bf K}{\bf r}}d^3x=\sum\limits_{j=1}^3
\int\Psi_j^{\dag}\Psi_j e^{-i{\bf K}{\bf r}}d^3x =\sum\limits_{j=1}^3f_j(K)=$$ 
$$\sum\limits_{j=1}^3
\int (|F_{j1}|^2+|F_{j2}|^2)\Phi_j^{\dag}\Phi_j
\frac{e^{-i{\bf K}{\bf r}}}{r^2}d^3x,\>
d^3x=r^2\sin{\vartheta}dr d\vartheta d\varphi\eqno(14)$$
with the momentum transfer $K$. At $n_j=0=l_j$, as is easily seen, for any  
spinor of (11) we have $\Phi_j^{\dag}\Phi_j=1/(4\pi)$, so the integrand in 
(14) does not depend on $\varphi$ and we can consider vector ${\bf K}$ to be 
directed along z-axis. Then ${\bf Kr}=Kr\cos{\vartheta}$ and with the help of 
(9) and relations (see Ref. \cite{PBM1}): $\int_0^\infty 
r^{\alpha-1}e^{-pr}dr=
\Gamma(\alpha)p^{-\alpha}$, Re $\alpha,p >0$, 
$\int_0^\infty r^{\alpha-1}e^{-pr}\pmatrix{\sin{(Kr)}\cr\cos{(Kr)}\cr}dr=
\Gamma(\alpha)(K^2+p^2)^{-\alpha/2}
\pmatrix{\sin{(\alpha\arctan{(K/p))}}\cr\cos{(\alpha\arctan{(K/p))}}\cr}$, 
Re $\alpha >-1$, 
Re $p > |{\rm Im}\, K|$, $\Gamma(\alpha+1)=\alpha\Gamma(\alpha)$, 
$\int_0^\pi e^{-iKr\cos{\vartheta}}\sin{\vartheta}d\vartheta=2\sin{(Kr)}/(Kr)$, 
we shall obtain 
$$ f(K)=\sum\limits_{j=1}^3f_j(K)=
\sum\limits_{j=1}^3\frac{(2\beta_j)^{2\alpha_j+1}}{6\alpha_j}\cdot
\frac{\sin{[2\alpha_j\arctan{(K/(2\beta_j))]}}}{K(K^2+4\beta_j^2)^{\alpha_j}}$$
$$=\sum\limits_{j=1}^3\left(\frac{1}{3}-\frac{2\alpha^2_j+3\alpha_j+1}
{6\beta_j^2}\cdot \frac{K^2}{6}\right)+O(K^4), \eqno(15)$$
wherefrom it is clear that $f(K)$ is a function of $K^2$, as should be, and 
we can determine the root-mean-square radius of meson (quarkonium) in the form 
$$<r>=\sqrt{\sum\limits_{j=1}^3\frac{2\alpha^2_j+3\alpha_j+1}
{6\beta_j^2}}.\eqno(16)$$
When calculating (15) also the fact was used that by virtue of the 
normalization condition for wave 
functions we have $C_j^2[P_j^2(1-gb_j/\beta_j)^2+Q_j^2(1+gb_j/\beta_j)^2]=
(2\beta_j)^{2\alpha_j+1}/[3\Gamma(2\alpha_j+1)]$.

It is clear, we can directly calculate $<r>$ in accordance with the standard 
quantum mechanics rules as $<r>=\sqrt{\int r^2\Psi^{\dag}\Psi d^3x}=
\sqrt{\sum\limits_{j=1}^3\int r^2\Psi^{\dag}_j\Psi_j d^3x}$ and the 
result will be the same as in (16). So we should not call $<r>$ of (16) 
the {\em charge} radius of meson (quarkonium)-- it is just the radius of meson 
(quarkonium) determined 
by the wave functions of (6) (at $n_j=0=l_j$) with respect to strong 
interaction, i.e., radius of confinement. 
Now we should note the expression (15) to depend on 3-vector ${\bf K}$. To 
rewrite it in the form holding true for any 4-vector $Q$, let us recall that 
according to general considerations (see, e.g., Ref. \cite{LL1}) the relation 
(15) should correspond to the so-called Breit frame where $Q^2=-K^2$ 
[when fixing the metric by (1)] so it is 
not complicated to rewrite (15) for arbitrary $Q$ in the form 
$$ f(Q^2)=\sum\limits_{j=1}^3f_j(Q^2)=
\sum\limits_{j=1}^3\frac{(2\beta_j)^{2\alpha_j+1}}{6\alpha_j}\cdot
\frac{\sin{[2\alpha_j\arctan{(\sqrt{|Q^2|}/(2\beta_j))]}}}
{\sqrt{|Q^2|}(4\beta_j^2-Q^2)^{\alpha_j}}\> \eqno(17) $$
which passes on to (15) in the Breit frame. 

\subsection{Magnetic moment}
We can define the volumetric magnetic moment density by 
${\bf m}=q({\bf r}\times {\bf J})/2=q[(yJ_z-zJ_y){\bf i}+
(zJ_x-xJ_z){\bf j}+(xJ_y-yJ_x){\bf k}]/2$ with the meson charge $q$ and 
${\bf J}=\Psi^{\dag}(I_3\otimes{\bf \alpha})\Psi$. Using (6) we have in the 
explicit form 
$$J_x=\sum\limits_{j=1}^3
(F^\ast_{j1}F_{j2}+F^\ast_{j2}F_{j1})\frac{\Phi_j^{\dag}\Phi_j}
{r^2},\> 
J_y=\sum\limits_{j=1}^3
(F^\ast_{j1}F_{j2}-F^\ast_{j2}F_{j1})
\frac{\Phi_j^{\dag}\sigma_2\sigma_1\Phi_j}{r^2},\>$$
$$J_z=\sum\limits_{j=1}^3
(F^\ast_{j1}F_{j2}-F^\ast_{j2}F_{j1})
\frac{\Phi_j^{\dag}\sigma_3\sigma_1\Phi_j}{r^2}  \eqno(18)$$
with Pauli matrices $\sigma_{1,2,3}$.
Magnetic moment of meson (quarkonium) is ${\bf M}=\int_V {\bf m}d^3x$, where 
$V$ is the volume 
of the meson (quarkonium) (the ball of radius $<r>$). Then at $n_j=l_j=0$, 
as is seen from (9), (11), $F^\ast_{j1}=F_{j1},F^\ast_{j2}=-F_{j2}$, 
$\Phi_j^{\dag}\sigma_2\sigma_1\Phi_j=0 $ for any spinor of (11) which entails 
$J_x=J_y=0$, i.e., $m_z=0$ while $\int_V m_{x,y}d^3x=0$ because of 
turning the integral over $\varphi$ to zero, which is easy to check.
As a result, the magnetic moments of mesons (quarkonia) with the 
wave functions of (6) (at $l_j=0$) are equal to zero, as should be according 
to experimental data \cite{pdg}. 

Though we can also evaluate the magnetic form factor $F(Q^2)$ of meson 
(quarkonium) which is also a function of $Q^2$ (see Refs. \cite{{Gon07a},{Gon07b}}) 
the latter will not be used in the given paper so we shall not dwell upon it. 

\section{An independent estimate of $<r>$}
Inasmuch as at present there exists no generally accepted estimate of $<r>$ 
for an $\eta^\prime$-meson \cite{pdg}, the question now is how to estimate 
$<r>$ independently and then calculate it within framework of our approach. For 
this aim, let us employ the present-day width of electromagnetic decay 
$\eta^\prime\to 2\gamma$ which is approximately 
equal to $\Gamma_5=4.3$ keV according to the notation of Ref. \cite{pdg}. 
In this situation, one can use a variant of formulae originating from 
Ref. \cite{Van67}. Such formulae 
are often employed, for example, in the heavy quarkonia physics (see, e. g., 
Ref. \cite{Bra05}). In their turn, they are actually based on the standard 
expression from the elementary kinetic theory of gases (see, e. g., 
Ref. \cite{{Sav89}}) for the number $\nu$ of collisions of a molecule per unit 
time
$$ \nu=\sqrt{2}\sigma <v>n\>,                  \eqno(19)$$  
where $\sigma$ is an effective cross section for molecules, $<v>$ is a mean 
molecular velocity, $n$ is the concentration of molecules. 
If replacing $\nu\to\Gamma_5$ we may fit (19) to estimate 
$\Gamma_5$ when interpreting $\sigma$ as the cross 
section of annihilation $\bar{q}q\to 2\gamma$ for the quark-antiquark pair 
due to electromagnetic interaction, $<v>$ and $n$ as, respectively, a mean quark 
velocity and the concentration of quarks (antiquarks) in meson (quarkonium). 
To gain $\sigma$ in the explicit form one may take the corresponding relation 
(Dirac formula) for the cross section of annihilation $e^+e^-\to 2\gamma$  
(see, e. g., Ref. \cite{{LL1}}) and, after replacing 
$\alpha_{em}\to Q\alpha_{em}$, $m_{e}\to m_q$ with electromagnetic coupling 
constant $\alpha_{em}$=1/137.0359895 and electron mass $m_{e}$, we obtain 
$$\sigma=\frac{N}{2}\pi r_q^2\frac{(\tau^2+\tau-\frac{1}{2})\ln{\frac{\sqrt{\tau}
+\sqrt{\tau-1}}{\sqrt{\tau}-\sqrt{\tau-1}}}-
(\tau+1)\sqrt{\tau(\tau-1)}}{\tau^2(\tau-1)}\eqno(20) $$
with $\tau=s/(4m_q^2)$, where the Mandelstam invariant 
$s=2m_q(m_q+\mu/2)$ with $\mu=957.78$ MeV, $Q=2/3$ for $\bar{u}u$-state of 
an $\eta^\prime$-meson while $Q=1/3$ for $\bar{d}d$- and $\bar{s}s$-states, 
$N$ is the number of colours, $r_q=\alpha_{em}Q^2/m_q$. 
To get $<v>$ one may use 
the standard relativistic relation $v=\sqrt{T(T+2E_0)}/(T+E_0)$ with kinetic 
$T$ and rest energies $E_0$ for velocity $v$ of a point-like particle. Putting 
$T=\mu/2-m_q$, $E_0=m_q$ we shall gain 
$$<v>=\sqrt{1-\frac{4m_q^2}{\mu^2}} \>.  \eqno(21)$$
At last, obviously, $n=1/V$, where the volume of meson (quarkonium) 
$V=4\pi<r>^3/3$ while $<r>$ may be calculated in accordance with (16). 
The relations (19)--(21) entail the sought independent estimate for $<r>$

$$<r>=\left(\frac{3\sigma\sqrt{2}\sqrt{1-\frac{4m_q^2}{\mu^2}}}
{4\pi\Gamma_5}\right)^{1/3}                  \eqno(22)$$
with $\sigma$ of (20). When inserting $N=3$, $\mu=957.78$ MeV, $m_q=2.25$ MeV, 
$5$ MeV, $107.5$ MeV, $\Gamma_5=4.3$ keV into (22) we shall have 
$<r>\approx19.127$ fm, 4.243 fm, 0.455 fm, respectively, 
for $\bar{u}u$-, $\bar{d}d$-, $\bar{s}s$-states of an $\eta^\prime$-meson. 
If noticing that for $\pi^0$-meson an experimental estimate of $<r>$ is 
of order 0.672 fm \cite{pdg}, for $K^0$-meson we have about 0.560 fm \cite{pdg} 
while for $\eta$-meson $<r>\approx0.54$ fm \cite{{Gon07b}} then it is 
reasonable to take $<r>\sim$ (0.40--0.45) fm for $\eta^\prime$-meson which is 
our choice within the present paper. At last, the corresponding quark 
velocities evaluated in accordance with (21) are $v_u=0.9999889856$, 
$v_d=0.9999456069$, $v_s=0.9745331899$ so quarks in an $\eta^\prime$-meson 
should be considered the ultrarelativistic point-like particles. 

\section{Estimates for parameters of SU(3)-gluonic field in 
$\eta^\prime$-meson}
\subsection{Basic equations and numerical results}
Now we are able to estimate parameters $a_j, b_j, B_j$ of the confining 
SU(3)-field (3) for an $\eta^\prime$-meson within framework of our approach. 
In this situation, we should consider (12) and (16) the system of equations 
which should be solved compatibly if taking $\mu= 957.78$ MeV, $m_u= 2.25$ MeV, 
$m_d= 5.0$ MeV, $m_s= 107.5$ MeV and 
$<r>\,\approx$ (0.40--0.45) fm (see Section 4).
While computing for distinctness we take all the eigenvalues $\lambda_j$ of the 
Euclidean Dirac operator ${\cal D}_0$ on the unit 2-sphere ${\mathbb S}^2$ 
equal to 1. The
results of numerical compatible solving of equations (12) and (16)
are adduced in Tables 1--2.

\begin{table}[htbp]
\caption{Gauge coupling constant, reduced mass $\mu_0$ and
parameters of the confining SU(3)-gluonic field for an $\eta^\prime$-meson}
\label{t.1}
\begin{center}
\begin{tabular}{|c|c|c|c|c|c|c|c|c|}
\hline
\small Particle & \small $ g$ & \small $\mu_0$ (\small MeV) & \small $a_1$ 
& \small $a_2$ & \small $b_1$ (\small GeV) & \small $b_2$ (\small GeV) 
& \small $B_1$ & \small $B_2$ \\
\hline
$\eta^\prime$---$\bar{u}u$  & \scriptsize 3.91476 & \scriptsize 1.125 & 
\scriptsize 0.218474 & \scriptsize -0.394718 & 
\scriptsize 0.618419 & \scriptsize -0.280807 & 
\scriptsize -0.300 & \scriptsize  -0.200 \\
\hline
$\eta^\prime$---$\bar{d}d$  & \scriptsize 3.91476 & \scriptsize 2.50 & 
\scriptsize 0.351384 & \scriptsize -0.130858 & 
\scriptsize 0.278983 & \scriptsize 0.285548 & 
\scriptsize -0.150 & \scriptsize  0.410 \\
\hline
$\eta^\prime$---$\bar{s}s$  & \scriptsize 3.91476 & \scriptsize 53.75 & 
\scriptsize 0.123645 & \scriptsize 0.124633 & 
\scriptsize -0.226875 & \scriptsize 0.588802 & 
\scriptsize 0.410 & \scriptsize  -0.160 \\
\hline
\end{tabular}
\end{center}
\end{table}

\begin{table}[htbp]
\caption{Theoretical and experimental $\eta^\prime$-meson mass and radius}
\label{t.2}
\begin{center}
\begin{tabular}{|c|c|c|c|c|} 
\hline
\tiny Particle & \tiny Theoret. $\mu$ (MeV) &  \tiny Experim. $\mu$ (MeV) & 
\tiny Theoret. $<r>$ (fm)  & \tiny Experim. $<r>$ (fm)  \\
\hline
\scriptsize $\eta^\prime$---$\bar{u}u$   & \scriptsize $\mu= 2m_u+
\omega_j(0,0,1)= 957.78$ & \scriptsize 957.78 & \scriptsize 0.404140 & 
\scriptsize - \\
\hline
\scriptsize $\eta^\prime$---$\bar{d}d$ & \scriptsize $\mu =2m_d+
\omega_j(0,0,1)= 957.78$ & \scriptsize 957.78 & \scriptsize 0.404382 & 
\scriptsize - \\                                           
\hline
\scriptsize $\eta^\prime$---$\bar{s}s$ & \scriptsize $\mu =2m_s+
\omega_j(0,0,1)= 957.78$ & \scriptsize 957.78 & \scriptsize 0.362994 & 
\scriptsize - \\                                           
\hline
\end{tabular}
\end{center}
\end{table}
\subsection{Consistency with the width of two-photon decay  
$\eta^\prime\to2\gamma$}
Let us consider whether the estimates of previous subsection are consistent 
with the width of the electromagnetic 2-photon decay  
$\eta^\prime\to2\gamma$.  
Actually kinematic analysis based on 
Lorentz- and gauge invariances gives rise to the following expression for 
the width $\Gamma$ of the electromagnetic decay $P\to2\gamma$ (where 
$P$ stands for any meson from 
$\pi^0$, $\eta$, $\eta^\prime$, see, e.g., Ref. \cite{RF})
$$ \Gamma=\frac{1}{4}\pi\alpha_{em}^2g^2_{P\gamma\gamma}\mu^3 \eqno(23) $$
with the electromagnetic coupling constant $\alpha_{em}$=1/137.0359895 and 
the $P$-meson mass $\mu$ while the information about strong 
interaction of quarks in $P$-meson is encoded in a decay constant 
$g_{P\gamma\gamma}$. Making replacement $g_{P\gamma\gamma}=
f_P/\mu $ we can reduce (23) to the form 
$$ \Gamma=\frac{\pi\alpha_{em}^2\mu f_P^2}{4}\>. \eqno(24) $$
Now it should be noted that the only 
invariant which $f_P$ might depend on is $Q^2=\mu^2$, i. e. we should find 
such a function ${\cal F}(Q^2)$ for that ${\cal F}(Q^2=\mu^2)=f_P$ but  
${\cal F}(Q^2)$ cannot be computed by perturbative techniques. It is 
obvious from the physical point of view that ${\cal F}(Q^2)$ should be connected 
with the electromagnetic properties of $P$-meson. As we have seen in 
Section 3, there are at least two suitable functions for this aim -- electric 
and magnetic form factors. But there exist no experimental 
consequences related to a magnetic form factor at present whereas electric 
one to some extent determines, e. g., an effective size of meson (quarkonium) 
in the 
form $<r>$ of (16). It is reasonable, therefore, to take 
${\cal F}(Q^2=\mu^2)=Af(Q^2=\mu^2)$ with some constant $A$ and the electric form 
factor $f$ of (17) for the sought relation. In the situation, we obtain 
an additional equation imposed on parameters of the confining SU(3)-gluonic field 
in $P$-meson which has been used in Refs. \cite{{Gon07a},{Gon07b}} to 
estimate the mentioned parameters in $\pi^0$- and $\eta$-mesons. 
As a result, using (17) in the case of an $\eta^\prime$-meson, we come from 
(24) to relation 
$$ \Gamma=\frac{\pi\alpha_{em}^2\mu}{4}
\left(A\sum\limits_{j=1}^3\frac{1}{6\alpha_jx_j}\cdot
\frac{\sin{(2\alpha_j\arctan{x_j})}}
{(1-x_j^2)^{\alpha_j}}\right)^2
\approx4.3\> {\rm keV}\> \eqno(25) $$
with $x_j=\mu/(2\beta_j)$, $\mu=957.78$ MeV and we used the width  
$\Gamma=\Gamma_5\approx4.3$ keV 
for the electromagnetic decay $\eta^\prime\to2\gamma$ following the notation 
from Ref. \cite{pdg}. 
In the circumstances, we can employ the results of Table 1 and compute the 
left-hand side of (25) which entails 
the corresponding value $A\approx0.231$ for any state of 
$\eta^\prime$---$\bar{u}u$, $\eta^\prime$---$\bar{d}d$, 
$\eta^\prime$---$\bar{s}s$. Consequently, 
we draw the conclusion that parameters of the confining SU(3)-gluonic field 
in an $\eta^\prime$-meson from Table 1 might be consistent with $\Gamma_5$. 

\section{Estimates of gluon concentrations, electric and magnetic colour field 
strengths}
Now let us recall that, according to Refs. \cite{{Gon052},{Gon06}}, one can 
confront the field (3) with the $T_{00}$-component (the volumetric energy 
density of the SU(3)-gluonic field) of the energy-momentum tensor (2) so that 
$$T_{00}\equiv T_{tt}=\frac{E^2+H^2}{2}=\frac{1}{2}\left(\frac{a_1^2+
a_1a_2+a_2^2}{r^4}+\frac{b_1^2+b_1b_2+b_2^2}{r^2\sin^2{\vartheta}}\right)
\equiv\frac{{\cal A}}{r^4}+
\frac{{\cal B}}{r^2\sin^2{\vartheta}}\>\eqno(26)$$
with electric $E$ and magnetic $H$ colour field strengths and with 
real ${\cal A}>0$, ${\cal B}>0$. One can also introduce magnetic colour 
induction $B=(4\pi\times10^{-7} {\rm H/m})\,H$, where $H$ in A/m.  

To estimate the gluon concentrations
we can employ (26) and, taking the quantity
$\omega= \mit\Gamma$, the full decay width of a meson, for 
the characteristic frequency of gluons we obtain
the sought characteristic concentration $n$ in the form
$$n=\frac{T_{00}}{\mit\Gamma}\>, \eqno(27)$$
so we can rewrite (26) in the form 
$T_{00}=T_{00}^{\rm coul}+T_{00}^{\rm lin}$ conforming to the contributions 
from the Coulomb and linear parts of the
solution (3). This entails the corresponding split of $n$ from (27) as 
$n=n_{\rm coul} + n_{\rm lin}$. 

The parameters of Table 1 were employed when computing and for simplicity 
we put $\sin{\vartheta}=1$ in (26). There was also used the following 
present-day full decay width of an $\eta^\prime$-meson 
${\mit\Gamma}=0.203$ MeV, whereas the Bohr radius 
$a_0=0.529177249\cdot10^{5}\ {\rm fm}$ \cite{pdg}. 

Table 3 contains the numerical results for $n_{\rm coul}$, $n_{\rm lin}$, $n$, 
$E$, $H$, $B$ for the meson under discussion.
\begin{table}[htbp]
\caption{Gluon concentrations, electric and magnetic colour field strengths in 
an $\eta^\prime$-meson}
\label{t.4}
\begin{center}
\begin{tabular}{|lllllll|}
\hline
\scriptsize $\eta^\prime$---$\bar{u}u$: & \scriptsize 
$r_0=<r>= 0.404140 \ {\rm fm}$ & & &  & & \\
\hline 
\tiny $r$ (fm)& \tiny $n_{\rm coul}$ $ ({\rm m}^{-3}) $ & \tiny $n_{\rm lin}$ 
$ ({\rm m}^{-3}) $& \tiny $n$ (${\rm m}^{-3}) $ & \tiny $E$ $({\rm V/m})$ 
& \tiny $H$ $({\rm A/m})$ & \tiny $B$ $({\rm T})$\\
\hline
\tiny $0.1r_0$ & \tiny $ 0.140954\times10^{56}$   
& \tiny $ 0.564544\times10^{53}$ 
& \tiny $ 0.141518\times10^{56}$ 
& \tiny $ 0.353468\times10^{25}$  
& \tiny $ 0.300915\times10^{22}$ 
& \tiny $ 0.378141\times10^{16}$ \\
\hline
\tiny$r_0$ & \tiny$ 0.140954\times10^{52}$ 
& \tiny$ 0.564544\times10^{51}$ 
& \tiny$ 0.197408\times10^{52}$ 
& \tiny$ 0.353468\times10^{23}$  
& \tiny$ 0.300915\times10^{21}$  
& \tiny$ 0.378141\times10^{15}$ \\
\hline
\tiny$1.0$ & \tiny$ 0.376014\times10^{50}$  
& \tiny$ 0.922064\times10^{50}$ 
& \tiny$ 0.129808\times10^{51}$ 
& \tiny$ 0.577316\times10^{22}$  
& \tiny$ 0.121612\times10^{21}$  
& \tiny$ 0.152822\times10^{15}$ \\
\hline
\tiny$10r_0$ & \tiny$0.140954\times10^{48}$  
& \tiny$0.564544\times10^{49}$ 
& \tiny$0.578639\times10^{49}$ 
& \tiny$0.353468\times10^{21}$  
& \tiny$0.300915\times10^{20}$ 
& \tiny$0.378141\times10^{14}$ \\
\hline
\tiny$a_0$ & \tiny$0.479511\times10^{31}$  
& \tiny$0.329275\times10^{41}$ 
& \tiny$0.329275\times10^{41}$ 
& \tiny$0.206163\times10^{13}$ 
& \tiny$0.229813\times10^{16}$  
& \tiny$0.288791\times10^{10}$ \\
\hline
\scriptsize $\eta^\prime$---$\bar{d}d$: & \scriptsize 
$r_0=<r>= 0.404382\ {\rm fm}$  &  & &  & & \\
\hline 
\tiny $r$ (fm)& \tiny $n_{\rm coul}$ $ ({\rm m}^{-3}) $ & \tiny $n_{\rm lin}$ 
$ ({\rm m}^{-3}) $& \tiny $n$ (${\rm m}^{-3}) $ & \tiny $E$ $({\rm V/m})$ 
& \tiny $H$ $({\rm A/m})$ & \tiny $B$ $({\rm T})$\\
\hline
\tiny $0.1r_0$ & \tiny $0.113422\times10^{56}$   
& \tiny $0.468584\times10^{53}$ 
& \tiny $0.113891\times10^{56}$ 
& \tiny $0.317074\times10^{25}$  
& \tiny $0.274150\times10^{22}$ 
& \tiny $0.344507\times10^{16}$ \\
\hline
\tiny$r_0$ & \tiny$0.113422\times10^{52}$ 
& \tiny$0.468584\times10^{51}$ 
& \tiny$0.160281\times10^{52}$ 
& \tiny$0.317074\times10^{23}$  
& \tiny$0.274150\times10^{21}$  
& \tiny$0.344507\times10^{15}$ \\
\hline
\tiny$1.0$ & \tiny$0.303296\times10^{50}$  
& \tiny$0.766251\times10^{50}$ 
& \tiny$0.106955\times10^{51}$ 
& \tiny$0.518495\times10^{22}$  
& \tiny$0.110861\times10^{21}$  
& \tiny$0.139313\times10^{15}$\\
\hline
\tiny$10r_0$ & \tiny$0.113422\times10^{48}$  
& \tiny$0.468584\times10^{49}$ 
& \tiny$0.479926\times10^{49}$ 
& \tiny$0.317074\times10^{21}$  
& \tiny$0.274150\times10^{20}$ 
& \tiny$0.344507\times10^{14}$ \\
\hline
\tiny$a_0$ & \tiny$0.386778\times10^{31}$  
& \tiny$0.273633\times10^{41}$ 
& \tiny$0.273633\times10^{41}$ 
& \tiny$0.185158\times10^{13}$ 
& \tiny$0.209498\times10^{16}$  
& \tiny$0.263262\times10^{10}$\\
\hline
\scriptsize $\eta^\prime$---$\bar{s}s$: & \scriptsize 
$r_0=<r>= 0.362994\ {\rm fm}$  &  & &  & & \\
\hline
\tiny $r$ (fm)& \tiny $n_{\rm coul}$ $ ({\rm m}^{-3}) $ & \tiny $n_{\rm lin}$ 
$ ({\rm m}^{-3}) $& \tiny $n$ (${\rm m}^{-3}) $ & \tiny $E$ $({\rm V/m})$ 
& \tiny $H$ $({\rm A/m})$ & \tiny $B$ $({\rm T})$\\
\hline
\tiny $0.1r_0$ & \tiny $0.853605\times10^{55}$   
& \tiny $0.643673\times10^{53}$ 
& \tiny $0.860042\times10^{55}$ 
& \tiny $0.275068\times10^{25}$  
& \tiny $0.321312\times10^{22}$ 
& \tiny $0.403773\times10^{16}$ \\
\hline
\tiny$r_0$ & \tiny$0.853605\times10^{51}$ 
& \tiny$0.643673\times10^{51}$ 
& \tiny$0.149728\times10^{52}$ 
& \tiny$0.275068\times10^{23}$  
& \tiny$0.321312\times10^{21}$  
& \tiny$0.403773\times10^{15}$ \\
\hline
\tiny$1.0$ & \tiny$0.148202\times10^{50}$  
& \tiny$0.848134\times10^{50}$ 
& \tiny$0.996336\times10^{50}$ 
& \tiny$0.362443\times10^{22}$  
& \tiny$0.116634\times10^{21}$  
& \tiny$0.146567\times10^{15}$\\
\hline
\tiny$10r_0$ & \tiny$0.853605\times10^{47}$  
& \tiny$0.643673\times10^{49}$ 
& \tiny$0.652209\times10^{49}$ 
& \tiny$0.275068\times10^{21}$  
& \tiny$0.321312\times10^{20}$ 
& \tiny$0.403773\times10^{14}$ \\
\hline
\tiny$a_0$ & \tiny$0.188995\times10^{31}$  
& \tiny$0.302874\times10^{41}$ 
& \tiny$0.302874\times10^{41}$ 
& \tiny$0.129431\times10^{13}$ 
& \tiny$0.220407\times10^{16}$  
& \tiny$0.276972\times10^{10}$\\
\hline
\end{tabular}
\end{center}
\end{table}

\section{Discussion and concluding remarks}
\subsection{Discussion}
 As is seen from Table 3, at the characteristic scales
of an $\eta^\prime$-meson the gluon concentrations are huge and the 
corresponding fields (electric and magnetic colour ones) can be considered 
to be the classical ones with enormous strenghts. The part $n_{\rm coul}$ of 
gluon concentration $n$ connected with the Coulomb electric colour field is 
decreasing faster than $n_{\rm lin}$, the part of $n$ related to the linear 
magnetic colour field, and at large distances $n_{\rm lin}$ becomes dominant. 
It should be emphasized that in fact the gluon concentrations are much 
greater than the estimates given in Table 3  
because the latter are the estimates for maximal possible gluon frequencies, 
i.e. for maximal possible gluon impulses (under the concrete situation of 
an $\eta^\prime$-meson). As was mentioned in Section 1, 
the overwhelming majority of gluons between quarks is soft, i. e., with 
frequencies much less than 0.203 MeV so the corresponding concentrations 
much greater than those  
in Table 3. The given picture is in concordance with the one obtained 
in Refs. \cite{{Gon03},{Gon06},{Gon07a},{Gon07b}}. 
As a result, the confinement mechanism developed in 
Refs. \cite{{Gon01},{Gon051},{Gon052}} is also confirmed by the considerations 
of the present paper. 

It should be noted, however, that our results are of a preliminary character 
which is readily apparent, for example, from that the current quark masses 
(as well as the gauge coupling constant $g$) used in computation are known 
only within the 
certain limits, and we can expect similar limits for the magnitudes 
discussed in the paper so it is neccesary for further specification of the 
parameters for the confining SU(3)-gluonic field 
in an $\eta^\prime$-meson which can be obtained, for instance, by calculating 
the width of decay $\eta^\prime\to2\pi^0+\eta$ 
with the help of wave functions of $\eta$- and $\pi^{0}$-mesons discussed 
in Refs. \cite{{Gon07a},{Gon07b}}. We hope to continue analysing 
the given problems elsewhere. 

\subsection{Concluding remarks}
Finally we should ask: how to correlate the results obtained above with the 
relation of SQM mentioned in Section 1 
$$\eta^\prime=\frac{1}{\sqrt{3}}(\bar{u}u+\bar{d}d+\bar{s}s)\>,\eqno(28)$$
representing $\eta^\prime$-meson as a superposition of 
three quarkonia. It seems to us one should apply the notions of the pure and 
mixed states elaborated in quantum mechanics (see, e.g., Ref. \cite{DB}). 
From this point view we may consider an $\eta^\prime$-meson the mixed state of 
three pure states while each pure state is realized with the probability equal 
to $(1/\sqrt{3})^2=1/3$. Then, for example, the root-mean-square radius of 
an $\eta^\prime$-meson should be evaluated according to 
$$<r>=\sqrt{\frac{1}{3}<r>_{\bar{u}u}^2+\frac{1}{3}<r>_{\bar{d}d}^2+
\frac{1}{3}<r>_{\bar{s}s}^2}\approx0.390871\>{\rm fm} \eqno(29)$$
with $<r>_{\bar{q}q}$ adduced in Table 2. Similar remarks obviously hold true 
for other physical quantities characterizing the meson under consideration. 

\appendix
\section{Eigenspinors of Euclidean Dirac operator on $\mathbb{S}^2$}
We here represent some results about eigenspinors of the Euclidean Dirac 
operator on two-sphere ${\mathbb S}^2$ employed in the main part of the paper. 

When separating variables in the Dirac equation (4) there naturally 
arises the Euclidean Dirac operator ${\cal D}_0$ on the unit two-dimensional 
sphere ${\mathbb S}^2$ and we should know its eigenvalues with the corresponding 
eigenspinors. Such a problem also arises in the black hole theory while 
describing the so-called twisted spinors on Schwarzschild and 
Reissner-Nordstr\"om black holes and it was analysed in 
Refs. \cite{{Gon052},{Gon99}}, so we can use the results obtained 
therein for our aims. Let us adduce the necessary relations. 

The eigenvalue equation for
corresponding spinors $\Phi$ may look as follows
$${\cal D}_0\Phi=\lambda\Phi.\>\eqno(A.1)$$

As was discussed in Refs. \cite{Gon99}, the natural form of ${\cal D}_0$ in 
local coordinates $\vartheta, \varphi$ on the unit sphere ${\mathbb S}^2$ looks 
as 
$${\cal D}_0=-i\sigma_1\left[
i\sigma_2\partial_\vartheta+i\sigma_3\frac{1}{\sin{\vartheta}}
\left(\partial_\varphi-\frac{1}{2}\sigma_2\sigma_3\cos{\vartheta}
\right)\right]=$$
$$\sigma_1\sigma_2\partial_\vartheta+\frac{1}{\sin\vartheta}
\sigma_1\sigma_3\partial_\varphi- \frac{\cot\vartheta}{2}
\sigma_1\sigma_2         \eqno(A.2)$$
with the ordinary Pauli matrices
$$\sigma_1=\pmatrix{0&1\cr 1&0\cr}\,,\sigma_2=\pmatrix{0&-i\cr i&0\cr}\,,
\sigma_3=\pmatrix{1&0\cr 0&-1\cr}\,, $$
so that $\sigma_1{\cal D}_0=-{\cal D}_0\sigma_1$.

The equation $(A.1)$ was explored in Refs. \cite{Gon99}.
Spectrum of $D_0$ consists of the numbers
$\lambda=\pm(l+1)$              
with multiplicity $2(l+1)$ of each one, where $l=0,1,2,...$. Let us 
introduce the number $m$ such that $-l\le m\le l+1$ and the corresponding 
number $m'=m-1/2$ so $|m'|\le l+1/2$. Then the conforming eigenspinors of  
operator ${\cal D}_0$ are 
$$\Phi=\pmatrix{\Phi_1\cr\Phi_2\cr}= 
\Phi_{\mp\lambda}=\frac{C}{2}\pmatrix{P^k_{m'-1/2}\pm P^k_{m'1/2}\cr
P^k_{m'-1/2}\mp P^k_{m'1/2}\cr}e^{-im'\varphi}\> \eqno(A.3) $$
with the coefficient $C=\sqrt{\frac{l+1}{2\pi}}$ and $k=l+1/2$.  
These spinors form an orthonormal basis in $L_2^2({\mathbb S}^2)$ 
and are subject 
to the normalization condition
$$\int_{{\mathbb S}^2}\Phi^{\dag}\Phi d\Omega=
\int\limits_0^\pi\,\int\limits_0^{2\pi}(|\Phi_{1}|^2+|\Phi_{2}|^2)
\sin\vartheta d\vartheta d\varphi=1\>. \eqno(A.4)$$
Further, owing to the relation $\sigma_1{\cal D}_0=-{\cal D}_0\sigma_1$ we, 
obviously, have
$$ \sigma_1\Phi_{\mp\lambda}=\Phi_{\pm\lambda}\,.  \eqno(A.5)$$

As to functions $P^k_{m'n'}(\cos\vartheta)\equiv P^k_{m',\,n'}(\cos\vartheta)$ 
then they can be chosen by 
miscellaneous ways, for instance, as follows (see, e. g.,
Ref. \cite{Vil91})
$$P^k_{m'n'}(\cos\vartheta)=i^{-m'-n'}
\sqrt{\frac{(k-m')!(k-n')!}{(k+m')!(k+n')!}}
\left(\frac{1+\cos{\vartheta}}{1-\cos{\vartheta}}\right)^{\frac{m'+n'}{2}}\,
\times$$
$$\times\sum\limits_{j={\rm{max}}(m',n')}^k
\frac{(k+j)!i^{2j}}{(k-j)!(j-m')!(j-n')!}
\left(\frac{1-\cos{\vartheta}}{2}\right)^j \eqno(A.6)$$
with the orthogonality relation at $m',n'$ fixed
$$\int\limits_0^\pi\,{P^{*k}_{m'n'}}(\cos\vartheta)
P^{k'}_{m'n'}(\cos\vartheta)
\sin\vartheta d\vartheta={2\over2k+1}\delta_{kk'}
\>.\eqno(A.7)$$
It should be noted that square of 
${\cal D}_0$ is 
$${\cal D}^2_0=-\Delta_{{\mathbb S}^2}I_2+
\sigma_2\sigma_3\frac{\cos{\vartheta}}{\sin^2{\vartheta}}\partial_\varphi
+\frac{1}{4\sin^2{\vartheta}} +\frac{1}{4}\>,
\eqno(A.8)$$
while Laplacian on the unit sphere is
$$\Delta_{{\mathbb S}^2}=
\frac{1}{\sin{\vartheta}}\partial_\vartheta\sin{\vartheta}\partial_\vartheta+
\frac{1}{\sin^2{\vartheta}}\partial^2_\varphi=
\partial^2_\vartheta+\cot{\vartheta}\partial_\vartheta
+\frac{1}{\sin^2{\vartheta}}\partial^2_\varphi\>,
\eqno(A.9)$$
so the relation $(A.8)$ is a particular case of the so-called 
Weitzenb{\"o}ck-Lichnerowicz formulas (see Refs. \cite{81}). 
Then from $(A.1)$ it follows 
${\cal D}^2_0\Phi=\lambda^2\Phi$ and, when using the ansatz  
$\Phi=P(\vartheta)e^{-im'\varphi}=\pmatrix{P_1\cr P_2\cr}e^{-im'\varphi}$, 
$P_{1,2}=P_{1,2}(\vartheta)$, the equation ${\cal D}^2_0\Phi=\lambda^2\Phi$ 
turns into 
$$\left(-\partial^2_\vartheta-\cot{\vartheta}\partial_\vartheta +
\frac{m'^2+\frac{1}{4}}{\sin^2{\vartheta}}+
\frac{m'\cos{\vartheta}}{\sin^2{\vartheta}}\sigma_1\right)P=$$
$$\left(\lambda^2-\frac{1}{4}\right)P\>,
\eqno(A.10)$$
wherefrom all the above results concerning spectrum of ${\cal D}_0$ can be 
derived \cite{Gon99}.

When calculating the functions $P^k_{m'n'}(\cos\vartheta)$ directly, to our 
mind, it is the most convenient to use the integral expression \cite{Vil91}

$$P^k_{m'n'}(\cos\vartheta)=\frac{1}{2\pi}
\sqrt{\frac{(k-m')!(k+m')!}{(k-n')!(k+n')!}}\>
\int_{0}^{2\pi}\left(e^{i\varphi/2}\cos{\frac{\vartheta}{2}}+
ie^{-i\varphi/2}\sin{\frac{\vartheta}{2}}\right)^{k-n'}\times$$
$$\left(ie^{i\varphi/2}\sin{\frac{\vartheta}{2}}+
e^{-i\varphi/2}\cos{\frac{\vartheta}{2}}\right)^{k+n'}e^{im'\varphi}d\varphi 
\eqno(A.11)$$
and the symmetry relations ($z=\cos{\vartheta}$) 
$$P^k_{m'n'}(z)=P^k_{n'm'}(z), \>P^k_{m',-n'}(z)=P^k_{-m',\,n'}(z), 
\>P^k_{m'n'}(z)=P^k_{-m',-n'}(z)\,,$$ 
$$P^k_{m'n'}(-z)=i^{2k-2m'-2n'}P^k_{m',-n'}(z)\>. \eqno(A.12)$$
In particular
$$P^{k}_{kk}(z)=
\cos^{2k}{(\vartheta/2)},  
P^{k}_{k,-k}(z)=i^{2k}\sin^{2k}{(\vartheta/2)},
P^{k}_{k0}(z)=\frac{i^{k}\sqrt{(2k)!}}{2^k k!}\sin^{k}{\vartheta}\,,$$
$$ P^{k}_{kn'}(z)=i^{k-n'}\sqrt{\frac{(2k)!}{(k-n')!(k+n')!}}
\sin^{k-n'}{(\vartheta/2)}\cos^{k+n'}{(\vartheta/2)}\>. \eqno(A.13)$$ 
\subsection*{Eigenspinors with $\lambda=\pm1,\,\pm2$}
If $\lambda=\pm(l+1)=\pm1$ then $l=0$ and from $(A.3)$ it follows that 
$k=l+1/2=1/2$, $|m'|\le1/2$ and we need the functions $P^{1/2}_{m',\pm1/2}$ 
that are easily evaluated with the help of $(A.11)$--$(A.13)$ so   
the eigenspinors for $\lambda=-1$ are 
$$\Phi=\frac{C}{2}\pmatrix{\cos{\frac{\vartheta}{2}}+
i\sin{\frac{\vartheta}{2}}\cr
\cos{\frac{\vartheta}{2}}-i\sin{\frac{\vartheta}{2}}\cr}e^{i\varphi/2}, 
\Phi=\frac{C}{2}\pmatrix{\cos{\frac{\vartheta}{2}}+
i\sin{\frac{\vartheta}{2}}\cr
-\cos{\frac{\vartheta}{2}}+i\sin{\frac{\vartheta}{2}}\cr}
e^{-i\varphi/2},\eqno(A.14)$$
while for $\lambda=1$ the conforming spinors are
$$\Phi=\frac{C}{2}\pmatrix{\cos{\frac{\vartheta}{2}}-
i\sin{\frac{\vartheta}{2}}\cr
\cos{\frac{\vartheta}{2}}+i\sin{\frac{\vartheta}{2}}\cr}e^{i\varphi/2}, 
\Phi=\frac{C}{2}\pmatrix{-\cos{\frac{\vartheta}{2}}+
i\sin{\frac{\vartheta}{2}}\cr
\cos{\frac{\vartheta}{2}}+i\sin{\frac{\vartheta}{2}}\cr}e^{-i\varphi/2}
\eqno(A.15) $$
with the coefficient $C=\sqrt{1/(2\pi)}$.

It is clear that $(A.14)$--$(A.15)$ can be rewritten in the form 
$$\lambda=-1: \Phi=\frac{C}{2}\pmatrix{e^{i\frac{\vartheta}{2}}
\cr e^{-i\frac{\vartheta}{2}}\cr}e^{i\varphi/2},\> {\rm or}\>\>
\Phi=\frac{C}{2}\pmatrix{e^{i\frac{\vartheta}{2}}\cr
-e^{-i\frac{\vartheta}{2}}\cr}e^{-i\varphi/2},$$
$$\lambda=1: \Phi=\frac{C}{2}\pmatrix{e^{-i\frac{\vartheta}{2}}\cr
e^{i\frac{\vartheta}{2}}\cr}e^{i\varphi/2}, \> {\rm or}\>\>
\Phi=\frac{C}{2}\pmatrix{-e^{-i\frac{\vartheta}{2}}\cr
e^{i\frac{\vartheta}{2}}\cr}e^{-i\varphi/2}\,, 
\eqno(A.16) $$
so the relation $(A.5)$ is easily verified at $\lambda=\pm1$. 

In studying vector mesons and excited states of heavy quarkonia eigenspinors 
with $\lambda=\pm2$ may also be useful. Then $k=l+1/2=3/2$, $|m'|\le3/2$ and we 
need the functions $P^{3/2}_{m',\pm1/2}$ 
that can be evaluated with the help of $(A.11)$--$(A.13)$. Computation gives 
rise to 
$$ P^{3/2}_{3/2,-1/2}=-\frac{\sqrt{3}}{2}\sin{\vartheta}
\sin{\frac{\vartheta}{2}}= P^{3/2}_{-3/2,1/2},\>$$   
$$P^{3/2}_{3/2,1/2}=i\frac{\sqrt{3}}{2}\sin{\vartheta}
\cos{\frac{\vartheta}{2}}= P^{3/2}_{-3/2,-1/2},\>$$
$$P^{3/2}_{1/2,-1/2}= -\frac{i}{4}\left(\sin{\frac{\vartheta}{2}}-
3\sin{\frac{3}{2}\vartheta}\right)=P^{3/2}_{-1/2,1/2},\>$$
$$P^{3/2}_{1/2,1/2}= \frac{1}{4}\left(\cos{\frac{\vartheta}{2}}+
3\cos{\frac{3}{2}\vartheta}\right)=P^{3/2}_{-1/2,-1/2},\>
\eqno(A.17) $$
and according to ($A.3$) this entails eigenspinors with $\lambda=2$ in the 
form
$$\frac{C}{2}i\frac{\sqrt{3}}{2}\sin{\vartheta}
\pmatrix{e^{-i\frac{\vartheta}{2}}\cr
e^{i\frac{\vartheta}{2}}\cr}e^{i3\varphi/2},\>
\frac{C}{8}\pmatrix{3e^{-i\frac{3\vartheta}{2}}+e^{i\frac{\vartheta}{2}}\cr
3e^{i\frac{3\vartheta}{2}}+e^{-i\frac{\vartheta}{2}}\cr}e^{i\varphi/2},\>$$
$$\frac{C}{8}\pmatrix{-3e^{-i\frac{3\vartheta}{2}}-e^{i\frac{\vartheta}{2}}\cr
3e^{i\frac{3\vartheta}{2}}+e^{-i\frac{\vartheta}{2}}\cr}e^{-i\varphi/2},\>
\frac{C}{2}i\frac{\sqrt{3}}{2}\sin{\vartheta}
\pmatrix{-e^{-i\frac{\vartheta}{2}}\cr
e^{i\frac{\vartheta}{2}}\cr}e^{-i3\varphi/2}\>
 \eqno(A.18) $$
with $C=1/\sqrt{\pi}$, while eigenspinors with $\lambda=-2$ are obtained in 
accordance with relation $(A.5)$. 

\section{Radial parts}
We here adduce the explicit form for the radial parts of meson wave functions 
from (6). At $n_j=0$ they are given by 
$$F_{j1}=C_jP_jr^{\alpha_j}e^{-\beta_jr}\left(1-
\frac{Y_j}{Z_j}\right),F_{j2}=iC_jQ_jr^{\alpha_j}e^{-\beta_jr}\left(1+
\frac{Y_j}{Z_j}\right),\eqno(B.1)$$
while at $n_j>0$ by
$$F_{j1}=C_jP_jr^{\alpha_j}e^{-\beta_jr}\left[\left(1-
\frac{Y_j}{Z_j}\right)L^{2\alpha_j}_{n_j}(r_j)+
\frac{P_jQ_j}{Z_j}r_jL^{2\alpha_j+1}_{n_j-1}(r_j)\right],$$
$$F_{j2}=iC_jQ_jr^{\alpha_j}e^{-\beta_jr}\left[\left(1+
\frac{Y_j}{Z_j}\right)L^{2\alpha_j}_{n_j}(r_j)-
\frac{P_jQ_j}{Z_j}r_jL^{2\alpha_j+1}_{n_j-1}(r_j)\right]\eqno(B.2)$$
with the Laguerre polynomials $L^\rho_{n}(r_j)$, $r_j=2\beta_jr$, 
$\beta_j=\sqrt{\mu_0^2-\omega_j^2+g^2b_j^2}$ at $j=1,2,3$ with 
$b_3=-(b_1+b_2)$, 
$P_j=gb_j+\beta_j$, $Q_j=\mu_0-\omega_j$,
$Y_j=P_jQ_j\alpha_j+(P^2_j-Q^2_j)ga_j/2$, 
$Z_j=P_jQ_j\Lambda_j+(P^2_j+Q^2_j)ga_j/2$    
with $a_3=-(a_1+a_2)$,   
$\Lambda_j=\lambda_j-gB_j$ with $B_3=-(B_1+B_2)$, 
$\alpha_j=\sqrt{\Lambda_j^2-g^2a_j^2}$, 
while $\lambda_j=\pm(l_j+1)$ are
the eigenvalues of Euclidean Dirac operator ${\cal D}_0$ 
on unit two-sphere with $l_j=0,1,2,...$ (see Appendix A) 
and quantum numbers $n_j=0,1,2,...$ are defined by the relations 
$$n_j=\frac{gb_jZ_j-\beta_jY_j}{\beta_jP_jQ_j}\,, 
\eqno(B.3)$$
which entails the quadratic equation (7) and spectrum (8).  
Further, $C_j$ of $(B.1)$--$(B.2)$ should be determined
from the normalization condition
$$\int_0^\infty(|F_{j1}|^2+|F_{j2}|^2)dr=\frac{1}{3}\>.\eqno(B.4)$$
As a consequence, we shall gain that in (6) 
$\Psi_j\in L_2^{4}({\mathbb R}^3)$ at any $t\in{\mathbb R}$ and, accordingly,
$\Psi=(\Psi_1,\Psi_2,\Psi_3)$ may describe relativistic bound states 
in the field (3) with the energy spectrum (8). As is clear from $(B.3)$ at 
$n_j=0$ we have 
$gb_j/\beta_j=Y_j/Z_j$ so the radial parts of $(B.1)$ can be rewritten as  
$$F_{j1}=C_jP_jr^{\alpha_j}e^{-\beta_jr}\left(1-
\frac{gb_j}{\beta_j}\right),F_{j2}=iC_jQ_jr^{\alpha_j}e^{-\beta_jr}\left(1+
\frac{gb_j}{\beta_j}\right)\>.\eqno(B.5)$$
More details can be found in Refs. \cite{{Gon01},{Gon052}}.

\end{document}